# Acoustic omni meta-atom for top-down, decoupled access to all octants of a wave parameter space

Sukmo Koo[1], Choonlae Cho[1], Jun-ho Jeong[2] and Namkyoo Park[1]★

[1]*Photonic Systems Laboratory, Department of Electrical and Computer Engineering, Seoul National University, Seoul, Korea 08826*

[2]*Department of Nano Manufacturing Technology, Korea Institute of Machinery and Materials, Daejeon, Korea 34103*

★*E-mail address for correspondence: nkpark@snu.ac.kr*

**The common behavior of a wave is determined by wave parameters of its medium, which are generally associated with the characteristic oscillations of its corresponding elementary particles. In the context of metamaterials, the decoupled excitation of these fundamental oscillations would provide an ideal platform for top-down and reconfigurable access to the entire space of constitutive wave parameters; however, this has remained as a conceivable problem that must be accomplished, after being pointed out by Pendry[1]. Here, by focusing on acoustic metamaterials, we achieve the decoupling of density $\rho$, modulus $B^{-1}$, and bianisotropy[2,3] $\xi$ near the Dirac point[4,5], by separating the paths of particle momentum to conform to the characteristic oscillations of each macroscopic wave parameter. Independent access to all octants of wave parameter space ($\rho$, $B^{-1}$, $\xi$) = (+/-,+/-,+/-) is thus realized using a single platform that we call an omni meta-atom; as a building block that achieves top-down access to the target properties of metamaterials. With precision access to the target ($\rho$, $B^{-1}$, $\xi$), we also propose a bianisotropic meta-surface for independent shaping of transmission- and reflection-wave fronts, and a zero-index bianisotropic waveguide for pressure-velocity conversion.**

The general features of wave propagation are ultimately determined by the properties of its medium, where the wave travels through. In order to achieve an extreme manipulation of wave propagation, the accessibility to the unusual space of wave parameters is therefore obligatory. A wide variety of extreme wave parameters and their applications have been realized for different waves and material systems in the context of metamaterials; spanning the fields of, acoustics[5-10], photonics[1-4,11-13], thermodynamics[14], elasto-dynamics[15], seismics[16], among others. Negative-[7,8], zero-[4,5], ultrahigh-index[11], hyperbolic-[17], anisotropic-[18], bianisotropic-[2,3], chiral-[13], and disordered- metamaterials[19] have been demonstrated, along with their applications towards cloaking[1,15,20], super-focusing[18], perfect absorption[9], iso-spectrality[19], meta-surface hologram[21], and frequency-agile memory[22].

With keen interest on applications, reconfigurable control of wave parameters has also become one of the main streams in wave physics[1,7,8,23,24]. Nonetheless, although the decoupling of fundamental wave parameters has been envisaged as an ideal platform toward the top-down and deterministic reconfiguration of the meta-atom (Pendry *et al.*[1]), its feasibility has remained merely as a plausible idea that has yet to be responded. In most cases, the decoupling of constituent parameters has been achieved via the combination of elementary resonators in a non-isotropic and polarization-dependent form. As well, at present, strategies for metamaterial design have been based on bottom-up approaches; in which building blocks are proposed first, and subsequent design is performed iteratively until a specific index and impedance are achieved. Although it has been recently reported that pentamode metamaterials can provide all positive mechanical wave parameters[25], however, how to achieve full accessibility to the entire space of wave parameters with the designs of existing metamaterials remains an open question, and the existence of an omnipotent meta-atom platform for reconfigurable and seamless access to the wave parameter space also has yet to be answered.

Inspired by the fundamental oscillations of the elementary particle of a wave, in this communication we propose an entirely new design strategy for the meta-atom. Focusing on acoustic platform, the criteria for the decoupling of wave parameters are derived from first principles, and an omni meta-atom that achieves independent, full access to all octants of the wave parameter space ($\rho$, $B^{-1}$, ξ) is demonstrated. Based on the top-down access capability of the meta-atom for target ($\rho$, $B^{-1}$, ξ), we then demonstrate a new class of meta-devices; bianisotropic meta-surfaces for independent beam shaping of transmission- and reflection-waves, and as well as zero-index waveguides for pressure-velocity conversion. Our work provides a deeper insight on the relationship between wave parameters and the internal structures of a meta-atom, and paves a new route toward systematic access to target wave parameters.

Understanding that the electromagnetic wave parameters $\varepsilon$ and $\mu$ of a classical atom are directly related to the linear and angular oscillations of an electron, the insight of this study begins from the characteristic oscillation of elementary particles, in relation to wave parameters of interest. In this respect, the derivation of effective parameters for an acoustic wave ($\rho_x$: density for *x* direction, *B*: bulk modulus) from the characteristic motions of acoustic particles is straightforward (**Fig. 1a**).

Based on the duality between electromagnetic and acoustic waves[5], we first modify Alù's derivation of electromagnetic macroscopic wave parameters[12], to derive effective parameters of an acoustic system from first principles (details in **Supplement 1**). In the limit of a long wavelength (|**β**|*a* << $\pi$, **β**: effective wavevector, *a*: lattice constant), $B^{-1}$ and $\rho_x$ are then expressed as,

$$B^{-1} \sim \frac{\int_S B_s^{-1} p dS}{\int_S p dS + \frac{i\omega}{2}\int_S r(\rho_{sr}-1)v_r dS}, \quad \rho_x \sim \frac{\int_S \rho_{sx} v_x dS}{\int_S v_x dS - \frac{i\omega}{2}\int_S (B_s^{-1}-1)px dS} \quad (1)$$

for a two dimensional unit cell *S*; with a distributed particle density tensor $\rho_s$ (subscripts *x*, *r* denote density directions) and modulus $B_s$ (normalized to air) of the constituting materials inside *S*, where **r** is the position vector measured from the cell center, and *p* and **v** each correspond to the pressure and velocity fields at **r**.

Important to note from Eq. (1) is the presence of cross-coupling terms in the denominators of $\rho_x$ ($B_s^{-1}$) and $B^{-1}$($\rho_s$), which hinder the decoupled access to $\rho_x$ and $B^{-1}$. Out of various possibilities, we try to spatially decouple $\rho_x$ and $B^{-1}$ near $(\rho, B^{-1}) = (0, 0)$, by employing a meta-atom with an inner sub-cell (*IS*) of radial symmetry and outer sub-cells (*OS*) of linear vibrations conforming to the fundamental oscillations of $B^{-1}$ and $\rho_x$ (Fig. 1a) in a square lattice composed of a membrane, air, and solid walls (Fig. 1b).

Under these settings, the conditions of zero compressibility and zero density constrain the movement of outer and inner membranes, enabling further reduction of the equations; radial movement of the outer membrane is prohibited (as $B^{-1} \sim 0$), and outer- and inner-membrane should move out of phase (as $\rho_x \sim 0$) but with the same momentum value (details in **Supplement 2**). By employing a heavy mass for $\rho_m$ (or a large thickness for the inner membrane), we then achieve,

$$B^{-1} = \frac{\int_{S0} B_{s0}^{-1} p\, dS}{\frac{i\omega}{2} \int_{ISm} r\rho_{mr} v_r\, dS}, \quad \rho_x = \frac{\int_{ISm} \rho_{mx} v_x\, dS + \int_{OSm} \rho_{mx} v_x\, dS}{\int_{OS0} v_x\, dS}. \tag{2}$$

where subscript *m* and 0 denote the material (membrane and air) for the given physical quantities (e.g., $\rho$, $B^{-1}$) at *S*, *IS*, and *OS*. Eq. (2) shows the direct control of effective $B^{-1}$ with mass $\rho_m$ of the inner-membrane in the denominator, which justifies the proposed approach of dividing the meta-atom into the inner- and outer- sub-cells that correspond to the fundamental oscillations of $\rho_x$ and $B^{-1}$. With the inner membrane mass $\rho_m$ determined for $B^{-1} \sim 0$, then the control of effective $\rho_x$ with the tuning of only outer-membrane mass (i.e., second term $\rho_{mx}$ in the numerator of $\rho_x$) is consequently realized.

A more explicit solution for the structure shown in Fig. 1b can be obtained by using the coupled mode theory (CMT). Applying Newton's law to the membranes and Hooke's law to the air region, the decoupled relation for $\rho$ ($t_O$) and $B^{-1}$ ($t_I$) are again confirmed in the long-wavelength limit, as shown in Eq. (3) (derivation in **Supplement 3**):

$$(\omega^2 \rho_0)\rho = (t_O \rho_{Al} + \frac{\rho_0 s_O}{2a})\omega^2/a - \frac{2B_0}{s_O} - \frac{B_0 s_I}{s_O^2}$$

$$(\frac{B_0 a^2}{4 s_O^2 B_0^{-1}})B^{-1} = (t_I \rho_{Al} + \frac{\rho_0 s_O}{2a_{in}} + \frac{\rho_0 s_I}{4a_{in}})\omega^2/a_{in} - \frac{B_0}{s_O} - \frac{4B_0}{s_I}. \tag{3}$$

Worth to mention, with Eq. (3), it is also possible to achieve independent control of ($\rho$, $B^{-1}$) as a function of pressures (~ $B_0$) or volumes/areas (~ $s_O$, $s_I$) in sub-cells.

The membrane motion produced by the FEM in the meta-atom and the schematic of the membrane are shown in Figs. 2a and 2b. Experimental realizations of the meta-atom and membranes (Fig. 2c and d) are shown in Figs. 2c and 2d. Details of the structure and material parameters are described in the Methods section. Using the exact solutions (S14), in Fig. 2e we visualize the mapping of ($\rho$, $B^{-1}$) in terms of membrane thickness of outer- and inner- sub-cell ($t_O$, $t_I$) at 1,300 Hz. From the plot, perfectly orthogonal decoupling between $\rho$ and $B^{-1}$ is observed, especially near ($\rho$, $B^{-1}$) = (0, 0), which is in excellent agreement with FEM and experimental results (Figs. 2f and g; see field patterns of ($\rho$, $B^{-1}$) modes in **Supplement 4**). It is important to note that, inverse determination of the meta-atom structural parameter ($t_O$, $t_I$) is also possible from the target ($\rho$, $B^{-1}$) values using Eq. (3) or Fig. 2e.

Extending the discussion beyond ($\rho$, $B^{-1}$), the proposed approach could be generalized to the other wave parameter axis of bianisotropy[2,3]. Meanwhile the bianisotropy $\xi$ has been demonstrated using $\Omega$-type metamaterials[2,3] in nano-photonics, yet need to be conceptualized and demonstrated for acoustic metamaterials. In parallel to electromagnetic bianisotropy that couples kinetic and potential energies (or equivalently, electric and magnetic fields), here we investigate the coupling constant $\xi$ that connects the velocity and the pressure field. Considering that $\xi$ is related to structural asymmetry[12], we choose to apply asymmetry in the form of $t_I \pm 1/2\Delta t$ to generate $\xi$ (Fig. 3a). The analytically derived $\xi$ (Eq. (S16) in **Supplement 5**; its approximation is shown in Eq. (4)) show a highly linear relation with $\Delta t$. Most importantly, near-perfect decoupling from ($\rho$, $B^{-1}$) near the Dirac point (Fig. 3c, d) is realized, in excellent agreement with the numerical and experimental results (Fig. 3b-d).

$$\xi = -\frac{Z_0 s_I \omega \rho_{Al}}{2 a a_{in}} \Delta t. \tag{4}$$

The salient feature of the bianisotropic medium is in the asymmetric impedance manipulation of the wave with exchange in kinetic and potential energy during wave propagation. Using the bianisotropic meta-atom at a matched zero index, here we report a perfect transmission between *different widths* (or *impedances*) of waveguides (Fig. 3e).

As shown in Fig. 3e, six meta-atoms in the output waveguide with non-zero $\xi$ and $\rho = B^{-1} = 0$ in addition to a layer of meta-atoms of $(\rho, B^{-1}, \xi) = (0, 0, 0)$ in the input side are used. The $\xi$ value for atoms in the output waveguide for complete impedance conversion is calculated from the ratio of input/output waveguide widths and the number of bianisotropic meta-atoms ($\xi = \log(w_1/w_2)/(2k_0 \cdot 6a)$; see **Supplement 6**). Achieving exact $(\rho, B^{-1}, \xi)$ values for the meta-atoms from the independent control of $(t_O, t_I, \Delta t)$, Fig. 3e shows the pressure field for super-focusing ($w_1/w_2 = 15$) calculated by the FEM. As shown in Fig. 3e, the exponential amplification of pressure in the bianisotropic meta-waveguides achieving ideal impedance conversion, and suppression of higher order mode excitation from the meta-atom array of matched zero index is clear. Excellent agreement with analytical ((S23), **Supplement 7**) and experimental results (Fig. 3f) for an extreme case of a *single* meta-atom for $(w_1/w_2) = 2$ are achieved.

As a final application example supported by the capability of precisely- and independently-addressing target $\rho$, $B^{-1}$, and also $\xi$ values, we demonstrate a bianisotropic wave front shaping in a meta-surface[26-29] context, of critical novelty in a transmission-reflection decoupled form. Under the notion of the generalized Snell's law[26], the transmission- and reflection-decoupled bianisotropic wave front shaping can be achieved only via *independent* control of $(\phi_R, \phi_T)$ at the meta-surface; where the controllability of $\xi$ for individual meta-atom plays a critical role in achieving $n_R \neq n_T$ while maintaining the same value of $n_{\text{eff}}$ over the entire surface.

To confirm the feasibility of the independent and arbitrary controllability of $(\phi_R, \phi_T)$ using the proposed meta-atom, in Fig. 4a we plot the phase shift contour $(\phi_R, \phi_T)$ in the parameter octant space of $(\rho, B^{-1}, \xi)$, achieving 50:50 power division for the transmitted and reflected waves (details in **Supplement 8**). It is again emphasized that, in the absence of bianisotropy ($\xi = 0$), it is impossible to adjust $(\phi_R, \phi_T)$ under the given 50:50 power splitting condition, as evident from Fig.4a. From target phase shifts $(\phi_R, \phi_T)$ of an individual meta-atom (in a 20 × 1 array, Fig. S7), calculations of $(\rho, B^{-1}, \xi)$ are obtained from Eq. (S24) or Fig. S6 (details in **Supplement 8**) which achieves ordinary $\Delta\phi(x) = 0$ or anomalous $\Delta\phi(x) \neq 0$ transmission and reflections. Subsequent, top-down determination of the corresponding ($t_O$, $t_I$, $\Delta t$) from ($\rho$, $B^{-1}$, $\xi$) is then straightforward. Independent control of the reflected wave, the transmitted wave, as well as the simultaneous control of the reflected- and transmitted- wave compared to the reference (left figure), corresponding to the phase maps at the bottom of the figure are shown in Fig. 4b.

Experimental realization of a bianisotropic meta-surface has also been carried out using a 10 × 1 meta-atom array, in a 70 × 150 × 7 cm box with an acoustic absorber and an 8 × 1 speaker array (Fig. 4c). With the finite dimension of the setup used in this study, experiment has been performed with an incidence wave normal to the meta-atom array. Figure 4d shows scattered pressure field patterns together with the reflection- and transmission- phase ($\phi_R$, $\phi_T$) of individual meta-atoms; dotted lines are from the target design, square marks are from the impedance tube measurements, and solid lines are from the pressure field scanning measurements. With precise access to ($\phi_R$, $\phi_T$) values from the control of ($\rho$, $B^{-1}$, $\xi$) in each meta-atom, decoupled manipulation of the reflection- and transmission- wave fronts are successfully achieved experimentally.

In summary, with the insight gained from fundamental oscillations of the wave supported by first principles of homogenization theory, we demonstrated an acoustic omni meta-atom that achieves decoupled access to the target wave-parameter in the octant space of ($\rho$, $B^{-1}$, $\xi$), with the tuning of structural factors of the meta-atom ($t_O$, $t_I$, $\Delta t$). Excellent agreements between CMT-based solutions, FEM-based numerical analysis and experiments have been observed, confirming the top-down design capability for an omni meta-atom that addresses target ($\rho$, $B^{-1}$, $\xi$) values. The feasibility of active tunability using pressures and volumes in sub-cells toward reconfigurable control of meta-atoms has also been confirmed. Using independent and deterministic control of wave parameters, novel applications of bianisotropic pressure-velocity impedance conversion, and reflection-transmission decoupled wave front shaping have been achieved. Our work opens a new paradigm in the design of meta-atoms by overcoming difficulties observed from the bottom-up approach, and provides an ideal platform by resolving the previously envisaged but unanswered issue of the decoupled excitation of constitutive parameters. Using the same approach, we expect further extension of decoupled access to other waves (i.e., electromagnetic, elastic and thermal) and wave parameters (i.e., stress, strain, gyrotropy and chirality).

**Methods**

For the experiment, we constructed a 2D slab meta-atom (height = 7 cm) using an Al sheet-loaded LLDPE membrane and a solid Al wall (thickness = 3 mm, $a$ = 6 cm, $a_{in}$ = 2 cm; see Fig. 1a). To achieve the decoupling of ($\rho$, $B^{-1}$) at 1300 Hz, the effective thicknesses ($t_O$, $t_I$) of the Al sheet have been controlled between 35-60 and 50-90 μm, respectively; same thicknesses used in CMT and FEM analyses. The densities[30] of air and Al are assumed to be 1.21 and 2,700 kg/m$^3$. The wave parameters of interest have been calculated by using $S$ parameters extracted from a 4-point measurement impedance tube. It is noted that we employed a composite membrane constructed with an Al-sheet mounted on top of a larger frame of an LLDPE film as shown in Fig. 2b. The Al-sheet has a much greater weight and stiffness compared to the LLDPE film, and provides a method of controlling the composite membrane mass with its thickness; independently of the stiffness of the composite membrane ($k^{-1}_{composite} \sim k^{-1}_{LLDPE} + k^{-1}_{Al}$) that is primarily determined by the properties of the LLDPE film (10 μm thick). For the fine tuning of the membrane's effective thickness (i.e., mass), we used 3-8 stacked layers of Al-sheets that each had a thickness of ~15 μm with periodically perforated disks of a 2-mm radius. The final dimensions of the outer (inner) LLDPE film and the Al-sheet were 54 mm × 60 mm (18 mm × 60 mm) and 52 mm × 58 mm (16 mm × 58 mm), respectively.

**Acknowledgments**

This study was supported by the National Research Foundation funded by the Ministry of Science, ICT and Future Planning (Global Frontier Program: No. NRF-2014M3A6B3063708 and Global Research Laboratory: No. K20815000003).

**Author Contributions**

S.K. and N.P. devised the structure of the meta-atom. S.K. performed the theoretical analysis of the meta-atom and wrote the manuscript. C.C. prepared the experimental set-up and performed measurements. J.J. fabricated the meta-atom structures. N.P. conceived the idea, checked the theoretical derivations and wrote the manuscript. All authors contributed to the revisions of the manuscript.

**Additional Information**

Supplementary information is available in the online version of this paper. Reprints and permissions information are available online at http://www.nature.com/reprints. Correspondence and requests for materials should be addressed to N.P.

**Competing Financial Interests**

The authors declare no competing financial interest.

**Figure Legends**

**Figure 1 | Characteristic oscillations of acoustic atoms and the decoupling of constitutive parameters. a**, Linear and radial characteristic oscillations of acoustic atoms for $\rho$ and $B^{-1}$, respectively. **b**, Schematic of the proposed meta-atom. blue and red: outer and inner membranes; black: solid wall; $t_{O(I)}$ : outer (inner) membrane thickness; *OS* and *IS*: outer and inner region in the meta-atom unit cell.

**Figure 2 | Experimental realization of the meta-atom and the decoupling of constituent parameters. a**, Deflection pattern of the membrane obtained from 3D FEM. **b**, Structure of the membrane. **c, d**, 3D experimental realization of the meta-atom and membrane. **e**, CMT. **f**, 2D FEM (finite element method, COMSOL). **g**, experimental results showing decoupled access to $\rho$ (blue lines) and $B^{-1}$ (red lines) with outer and inner membranes thicknesses $t_O$ and $t_I$, respectively. Grid spacing for $\rho$ and $B^{-1}$ is 0.2. The thickness of the Al wall in the FEM and experiment is set at 3 mm, and the frequency is 1300Hz. The lattice constant and height of the cell were 6 cm and 7 cm, respectively. In the experiment, measurements were taken with a membrane thickness resolution at 10 μm.

**Figure 3 | Implementation of the bianisotropy $\xi$ in an acoustic meta-atom, and perfect transmission between different impedances using bianisotropic meta-waveguide. a**, Schematic of the bianisotropic meta-atom with an asymmetric arrangement of the membrane thickness. **b**, Tuning of $\xi$ with $\Delta t$ from the experiment (square symbols), CMT (solid line), FEM (dotted lines), and the approximation (point symbols, Eq. (4)). **c, d,** Tuning of ($\rho$, $\xi$) and ($B^{-1}$, $\xi$) with $\Delta t$. **e**, FEM-calculated pressure field for the super-focusing ($w_1/w_2 = 15$, $\xi = -0.158$, $\Delta t = -18$ μm). The inset shows the transmittance for super-focusing and super-radiation. **f**, Experimentally obtained super-focusing with a single meta-atom at different $\Delta t$ (or bianisotropy $\xi$). $w_1/w_2 = 2$.

**Figure 4 | Transmission and reflection decoupled wave front shaping using a bianisotropic meta-surface. a**, Phase shift contour ($\phi_R$, $\phi_T$) in the parameter octant space of ($\rho$, $B^{-1}$, $\xi$) for a 50:50 power division for the transmitted and reflected waves. **b,** FEM calculated pressure field patterns for an

incidence wave from the bottom at 45°. Left to right: reference, shift in the *reflection*, *transmission,* and *reflection and transmission* wavefront. The overlaid far-field polar plots are calculated from the near-field data. Transmission and reflection phases of the meta-atom array (dotted lines are for the design; the solid lines are measured from the FEM calculations) are shown at the bottom of each figure. **c,** Top view of the experimental setup. The experiment was performed with a 10 × 1 meta-atom array in a 70 × 150 × 7 cm box with an acoustic absorber and an 8 × 1 speaker array. **d,** Experimentally measured scattered pressure field patterns and calculated far-field polar plots for a normal incidence wave from the bottom. Below the field patterns, the transmission- and reflection- phases of the meta-atom array are shown. Dotted lines for design, square marks from the 1D impedance tube measurements, and solid lines from the experimentally measured pressure field near the meta-surface.

**Figures**

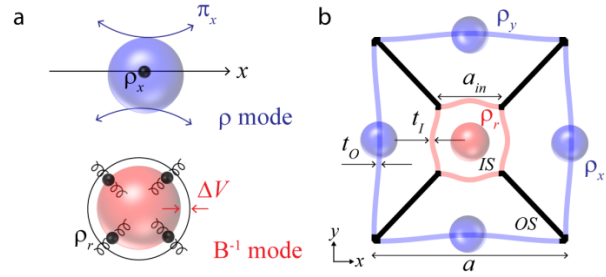

**Figure 1**

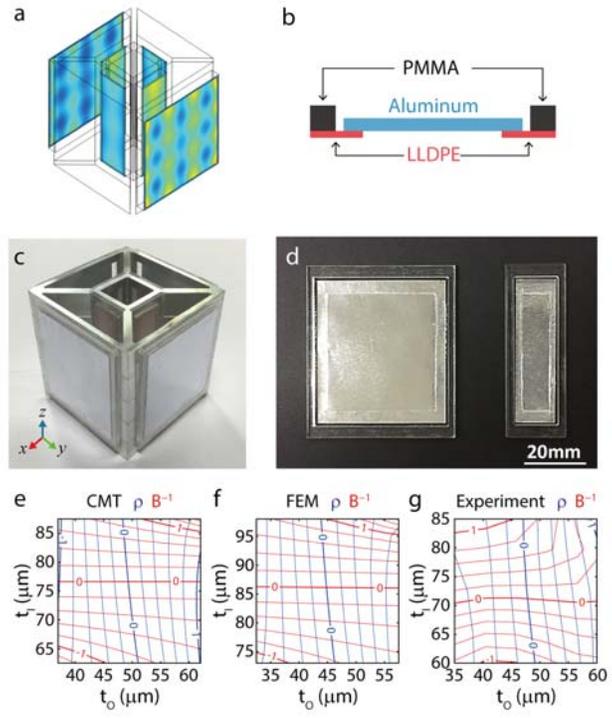

**Figure 2**

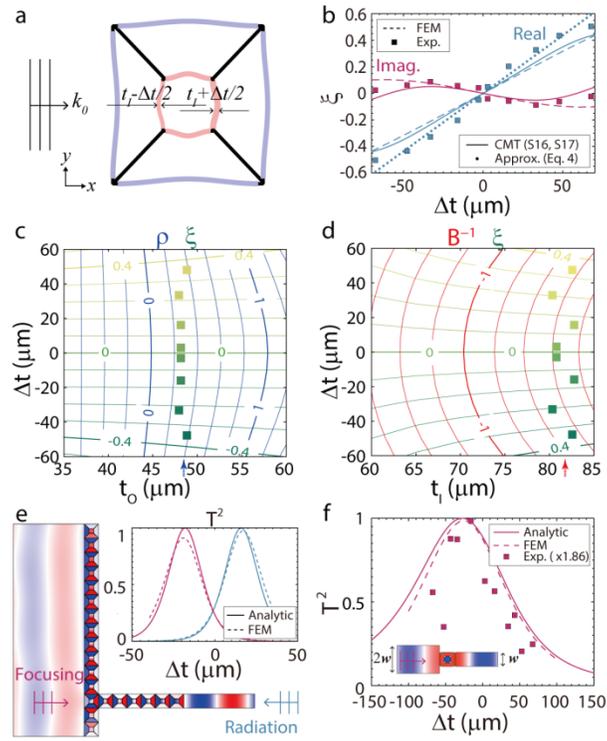

**Figure 3**

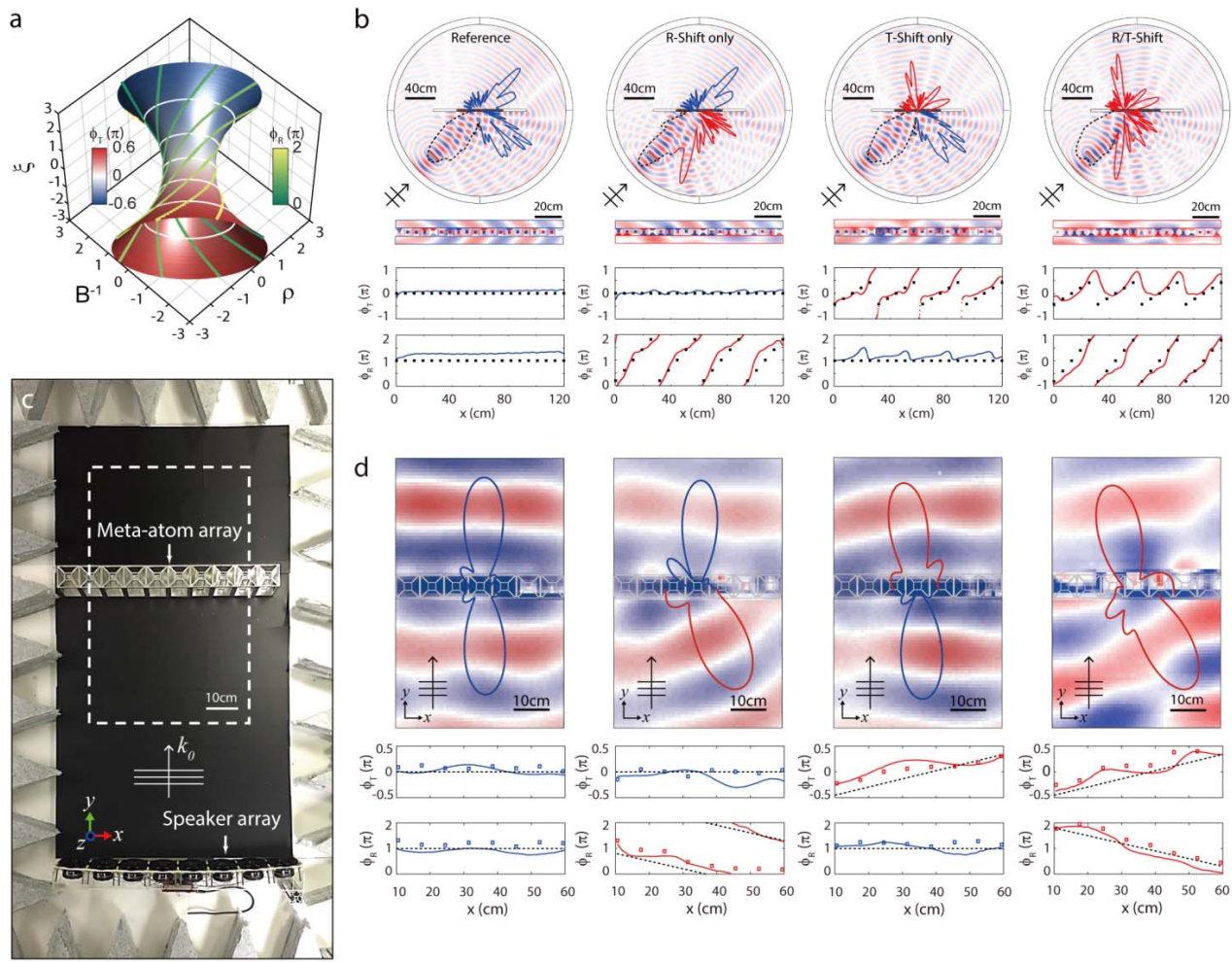

**Figure 4**